\documentclass[]{eptcs}
	\usepackage{breakurl}             
	\usepackage{underscore}           
	\usepackage{graphicx}
	\usepackage{times}

\usepackage{listings}
	\usepackage{float}
	\usepackage{adjustbox}
	\usepackage{graphicx}
	\usepackage{xcolor,colortbl}
	
	\definecolor{Gray}{gray}{0.85}
	\definecolor{LightCyan}{rgb}{0.88,1,1}
	
	\newcolumntype{a}{>{\columncolor{Gray}}c}
	\newcolumntype{b}{>{\columncolor{white}}c}

\title{Exploring the Link Between Test Suite Quality\\ and Automatic Specification Inference\thanks{Project GOMTA financed by the Malta Council for Science \& Technology through the National Research \&
		Innovation Programme 2013}}
	\author{Luke Chircop
	\institute{CS, ICT, University of Malta}
	\email{luke.chircop@um.edu.mt}
	\and
	Christian Colombo
	\institute{CS, ICT, University of Malta}
	\email{christian.colombo@um.edu.mt}
	\and
	Mark Micallef
	\institute{CS, ICT, University of Malta}
	\email{mark.micallef@um.edu.mt}
}
\def\titlerunning{Exploring the Link Between Test Suite Quality  and Automatic Specification Inference}
\def\authorrunning{L. Chircop, C. Colombo \& M. Micallef}

\begin{document}
	
	\maketitle

\begin{abstract}
While no one doubts the importance of correct and complete specifications, many industrial systems still do not have formal specifications written out --- and even when they do, it is hard to check their correctness and completeness. 
This work explores the possibility of using an invariant extraction tool such as Daikon to automatically infer specifications from available test suites with the idea of aiding software engineers to improve the specifications by having another version to compare to. 
Given that our initial experiments did not produce satisfactory results, in this paper we explore which test suite attributes influence the quality of the inferred specification. 
Following further study, we found that instruction, branch and method coverage are correlated to high recall values, reaching up to 97.93\%.

\end{abstract}

\section{Introduction}
\label{section:Introduction}

Having formally written specifications that are correct and complete is important. 
However, many industrial systems still do not have formal specifications written out. 
Even when they do, due to the size and complexity of such industrial systems, checking for correctness and completeness of the specification is rare.

Obtaining formally written specifications can be beneficial for multiple purposes. 
These can be used by a developer or tester as a point of reference to verify the implementation's behaviour. 
Formal specifications can also be exploited to aid test case generation \cite{MutTestGen} by pinpointing deficiencies in the automatically generated tests, mutation testing \cite{EffMutTestInvViol} by providing the means to identify valuable mutations minimising equivalent mutants, regression testing \cite{BCT} by pinpointing unintentional behavioural differences between different versions of the same system and to track down software bugs \cite{AutSoftFaultLocUGenProgInv, SoftBugDetAnomalyDet, FaultLocPotInv}. 

Instead of expecting companies or forcing developers to write formal specifications, there has been extensive research carried out \cite{ DaikonToAgitator, DynSyntProgInvUGenProg, DynDataBase, Daikon,AutGenOfProgSpec, EffIncAlgDynDetLikInv, StatSpecInf} for the possibility of automatically inferring specifications from readily available information. 
Apart from having to learn how to use the tool and prune some of the inferred specifications, acquiring the ability to infer specifications with minimal effort may encourage the industry to exploit the technology and use it to their advantage. 

Specifications can be inferred by using one or a combination of static and dynamic inference techniques. 
Static inference techniques process the source code and infer specifications based on the observations that are made. 
Dynamic inference techniques on the other hand, make use of drivers to exercise the system and specifications are inferred based on how the system is observed to react. 
A popular driver choice for such inference techniques are test suites. 

Given that around half of the development time for industrial systems is spent on testing, multiple test suites would have been written and be readily available. 
Furthermore, such test suites are purpose built to exercise as much of the system as possible, taking into consideration multiple input/output combinations to expose bugs in the system. 
This makes them an ideal source to use as drivers and gather knowledge with regards to how the system is expected to behave. 
Moreover, if used after all bugs identified (if any) are fixed, it can be safe to assume that the behaviour being observed is correct and will reduce the possibility of inferring incorrect specifications. 

One popular dynamic inference tool that has been used for various applications \cite{DaikonToAgitator, BCT, AutMutTestGenDynSymExec,EffIncAlgDynDetLikInv} is Daikon \cite{Daikon}. 
Daikon, is an invariant inference tool that makes use of test suites or other suitable resources to exercise systems and infer specifications based on what it observes. 
Since dynamic inference techniques infer program properties based on what is observed, the quality of the test suite greatly impacts the resulting correctness of the inferred specification. 
In fact, various testing attributes (used to measure the quality and effectiveness of written tests) including only testing for correct behaviour or covering all branch conditions could potentially have an effect on the outcome. 

In this paper, we make use of the Daikon tool to explore which test suite attributes influence the quality of the inferred specification. 
By doing so, this work will be able to provide clear and concise guidelines which may prove useful to users who want to infer and make use of specifications that correctly represent how a system should behave, by guiding them to focus on test suite attributes that were found to have the highest impact on specification quality. 
Moreover, such correlations can be used to predict the level of quality and correctness of the resulting specifications. 
Line and branch coverage are being hypothesised as potential test attributes that could have a noticeable affect on the quality of inferred specifications. 

In the rest of the paper, we present the state of the art in specification inference techniques and explore links between specification inference and test suite quality. 
More specifically, we start by introducing various approaches that exist to tackle this problem and discuss the reasoning behind the choice of the technique considered for the initial experiments in Section \ref{section:Background}. 
Subsequently, we introduce the case study (Section \ref{section:CaseStudy}) used throughout the evaluation process and report the empirical results in Section \ref{section:Evaluation}. 
Finally, we discuss related work (Section \ref{section:RelatedWork}), future work (Section \ref{section:FutureWork}) and conclude the paper (Section \ref{section:Conclusion}).

\section{Inferring Program Specifications} 
\label{section:Background}

The ultimate goal for any automatic specification inference technique is to be able to infer both complete and correct specifications on their own without any human intervention. 
However, it is both intractable and NP-hard as proven by Gold \cite{GOLD1978302} to have a program that determines on its own whether properties inferred, over or under generalises correct behaviour. 

Nonetheless, many techniques and tools have been introduced \cite{Daikon, SoftBugDetAnomalyDet, BCT, StatSpecInf} that automatically approximate program specifications based on information that is provided or observed. 
Two varying approaches divide the state of the art: static and dynamic inference techniques. 

\subsection{Static inference}
\label{subsection:StaticInference}

Static inference techniques infer program properties by examining the software's source code and understanding how variables and methods interact with each other. 
Ramanathan et al.\ \cite{StatSpecInf}, for example, introduced data flow analysis to infer control-flow and data-flow specifications able to represent the order in which methods can be executed and the state that variables have to be in upon or after execution of a method. 
Other works \cite{InferStatAnalysis,ESCJAVA}, make use of static inference techniques to discover common programming errors and validate program properties. 

This approach can be inefficient and may also require explicit goals or special annotations to be included in the source code; making it difficult and tedious at times \cite{AutGenOfProgSpec}. 
Furthermore, the technique is not able to reason about behaviour that is runtime dependent. 

\subsection{Dynamic inference}
\label{subsection:DynamicInference}

On the other hand, dynamic inference techniques \cite{DaikonToAgitator, DynDataBase, Daikon, BCT, AutGenOfProgSpec} infer program specifications by exercising the software and observing how it reacts to various inputs. 
Numerous sources can be used to exercise the software, however, test suites are most commonly used due to the implicit knowledge they hold. 
 
During the first phase of the approach, the software is exercised using the selected source driver and a trace is recorded. 
The recorded trace represents the program and data-flow of the exercised software. 
These traces are then processed in the second phase of the approach, whereby generalised specifications are inferred based on the behaviour that was observed. 
Program specifications have been modelled in multiple ways including: finite state machines using techniques such a k-tail \cite{Ktail} and invariants representing the manner in which methods can interact with each other and defining the state which variables should be in before or after the execution of methods, respectively. 

An assortment of tools exist that implement the dynamic inference approach including the Agitator tool \cite{DaikonToAgitator} that infers invariants using guided test input generation in an iterative process, the Behaviour, Capture and Test tool \cite{BCT} that exercises software and exploits the invariants generated to aid bug finding during regression testing and iDiscovery \cite{FeedDrivDynInvDisc} which combines dynamic inference and symbolic execution in an iterative feedback process to continuously refine the inferred specification until a fixed point is reached. 
All tools have one thing in common, that is, all make use of a tool called Daikon \cite{Daikon}.  

Daikon was first introduced by Michael D. Ernst et al.\ in 2000. 
It is able to infer invariants by instrumenting the software, exercising the instrumented software (e.g., using test suites), and finally inferring specifications based on the recorded traces. 
This was achieved by having the inference phase test the recorded values of instrumented variables against a predefined set of possible invariants \cite{Daikon, AutGenOfProgSpec}. 
Furthermore, some pruning and statistical techniques are implemented to reduce the number of undesirable invariants and present potentially relevant ones \cite{QuickDetRelProgInv}. 

As stated in \cite{AutGenOfProgSpec} and many other papers \cite{Daikon,CompStudyProgWrittenAutInfCont,FeedDrivDynInvDisc}, the accuracy and correctness of the specifications that are inferred by dynamic inference tools are in part dependent on the quality and effectiveness of the test suites used to exercise the software during the trace recording process. 
This paper makes use of the Daikon tool to explore which test suite attributes have a positive influence on the quality of the inferred specifications.

\section{Case Study}
\label{section:CaseStudy}

To be able to discover correlations between test suite attributes and their effect on the quality of inferred specifications, a financial transaction system (FTS in short) \footnote{The implementation is available at: https://github.com/lukechircop007/AutomaticSpecificationInferenceEvaluation.git} was designed and built. 
Included in the FTS case study is: (i) the software containing core functionality, (ii) a combination of four test suites with varying attributes testing the functionality of the software, and (iii) manually written specifications representing the desired behaviour of the software.  
\newline

\textbf{Software: } The core functionality of the FTS software was designed and built to replicate how a system would have been built in industry containing 1549 lines of code. 
Therefore, a multi level system with a facade implementation approach was taken. 
The API functionality accepts requests for: new user profiles to be created, accounts to be opened or closed and monetary transactions to and from local or foreign accounts to be performed, whilst also providing different user modes and account types that impact the manner in which some requests and charges are serviced. 
Moreover, the core functionality was also built to correctly handle invalid inputs or requests by throwing exceptions and returning pre-defined values, amongst others abilities. 
This was designed to showcase whether specification inference tools are able to infer specifications describing the handling of invalid requests or input. 
\newline 

\textbf{Test Suites: } Test suites were also built alongside the core functionality mimicking how the software would have been tested in industrial companies. 
To this end, unit tests (containing 437 lines of code) were introduced to separately validate the behaviour of low-level software functionality. 
Subsequently, integration tests (containing 1648 lines of code) were also introduced using a top-down approach to validate the high-level functionality that combined and interacted with various parts of the software. 
Test-to-pass (validating good behaviour) and test-to-fail (validating correct handling of invalid input or requests) tests were introduced at each level (marked as Valid and Invalid) to increase the total overall coverage of the software's behaviour. 
Boundary based testing techniques were also used whenever applicable to ensure coverage of corner cases. 

Every test was kept as simple as possible only exercising and checking for one property at a time with supporting set-up and tear-down functions. 
Furthermore, the inputs where pre-defined and fixed, facilitating repeatability of test runs (if bugs were to be uncovered).  
\newline 

\textbf{Specifications: } Finally, a specification was manually written representing how the software's functionality should behave given any input. 
This, provided the means to measure how much Daikon and other inference tools are able to infer correct and complete specifications by comparing the inferred specification with the manually written one and identifying matching, non matching and incorrect conditions. 
For this to be achieved, conditions of what input to accept and reactions to observe were defined for every method in the software. 
Some examples of such conditions include expecting an \textit{amount} value that is greater or equal to zero and expecting the \textit{balance} value to increase by \textit{amount} value after a successful \textit{deposit} request. 
The complete specification is provided within the FTS repository.

\section{Discovering Correlations between Specifications and Test Suites}
\label{section:Evaluation}

A selection of distinguishable test suite attributes and the resulting correctness of inferred specifications had to first be selected and evaluated in order to discover correlations between test suites and quality of specification. 
After obtaining the results, the study to discover correlations between test suites and specifications was conducted. 
Throughout the experiment, a developer machine consisting of an i7 quad core processor, eight gigabytes of RAM, running Windows 10 and the latest versions of Java (Java 8) was used. 

The rest of the section, discusses the design decisions involved (Sections \ref{subsection:ChoosingRelevantAttributes} and \ref{subsection:RunningExperiment}). 
Subsequently, we present and interpret the results and correlations that were identified in Sections \ref{subsubsection:CoverageMetricResults},  \ref{subsubsection:EvalCorrectnessSpec} and \ref{subsubsection:DiscAttQual}. 

\subsection{Choosing Relevant Test Suite Attributes}
\label{subsection:ChoosingRelevantAttributes}

Test suite attributes play an important role in the discovery of correlations since they are used to measure the quality and effectiveness of test suites allowing for one to be distinguishable from the other. 
A wide selection of metrics already exist that are able to showcase test suite attributes and have been used to evaluate how effective the written tests are at testing the software they were aimed for. 
Some metrics include measuring the percentage of software instructions exercised by the test suite and the percentage of conditional branch combinations covered. 

Rather than re-inventing the wheel, the same metrics were considered for the study. 
Further to the metrics mentioned above, line and method coverage were also taken into consideration. 
Since dynamic inference techniques infer specifications by analysing software behaviour, it stands to reason that having high software coverage levels will increase the possibility of discovering specifications that are of greater quality and completeness. 
In fact, we were hypothesising that having high branch and line coverage can positively influence the quality and correctness of the inferred specifications since the daikon tool is non-monotonic allowing conditions to be retracted based on further observed evidence. 

\subsection{Running the Experiment}
\label{subsection:RunningExperiment}

To determine the influence that test suite attributes had on the quality of inferred specifications, multiple test suites with varying metric values (some low, some high) were introduced resulting in a total combination of nine runs. 

\begin{figure}[h!]
	\centering
	\includegraphics[width=0.5\textwidth]{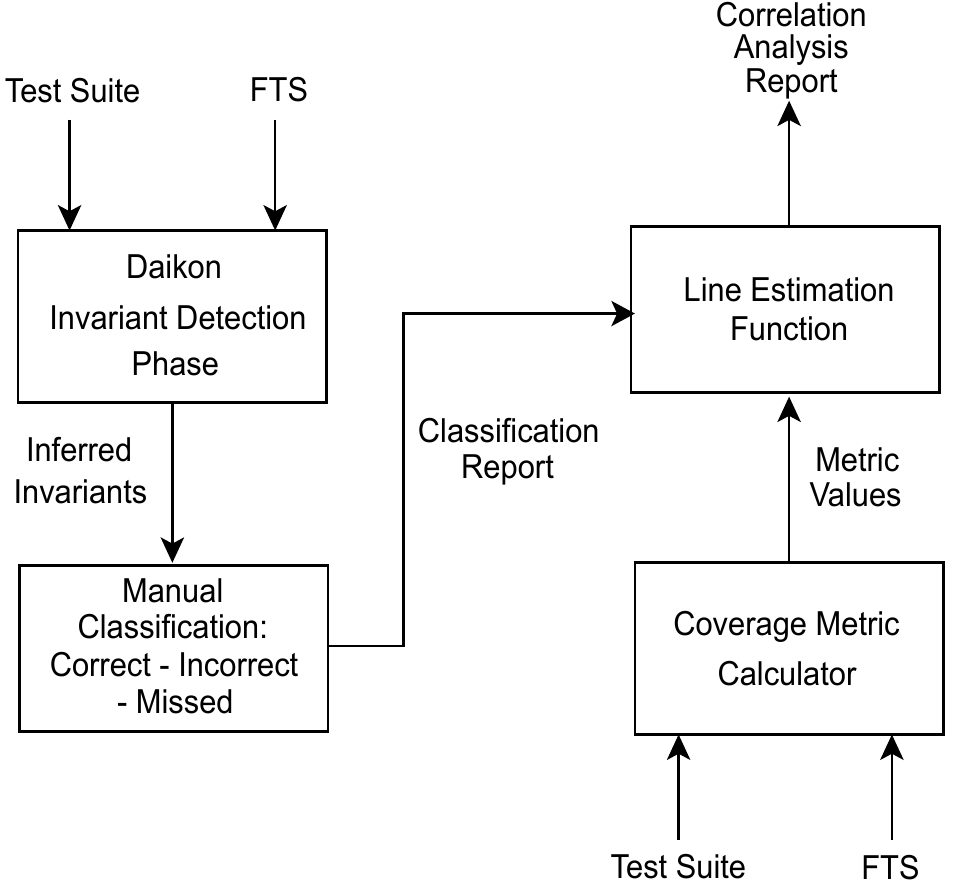}
	\caption{Evaluation process carried out to gather results for the discovery of potential correlations between test suite attributes and quality of inferred specification }
	\label{fig:Process}
\end{figure}

As shown in Figure \ref{fig:Process}, for each run:

\begin{itemize}
	\item Test suite metrics discussed in Section \ref{subsection:ChoosingRelevantAttributes} were measured. 
	\item Subsequently, Daikon was instructed to instrument a fresh version of the FTS and use the unique combination of tests defined for the particular run to infer a set of invariants.  
	\item Once inferred, the invariants were classified as correct or incorrect and a list of missing invariants was also included. 	
\end{itemize}

The results were then used to discover which test suite attributes positively influenced the quality of inferred specifications. 

\subsubsection{Coverage Metrics}
\label{subsubsection:CoverageMetricResults}

As mentioned in Section \ref{section:CaseStudy}, test-to-pass (marked as Valid) and test-to-fail (marked as Invalid) tests were introduced in both unit and integration tests. 
For the purpose of this study, a combination of valid, invalid, unit and integration tests was introduced for each run. 
When compared, the varying runs showcase distinctive coverage metric values with the last run encapsulating all tests, having the best coverage, as shown in Table \ref{table:Coverage Metrics}{}. 

\begin{table}[h!]
	\begin{center}
		\resizebox{\columnwidth}{!}{%
		\begin{tabular}{ |c||c|c|c|c|c|}
			\hline
			Test Suites Executed & Total Tests & Instruction (\%) & Branch (\%) & Line (\%) & Method (\%)\\
			\hline
			Valid: UT & 37 &	22.20 &	11.40 &	20.80 &	46.50\\		
			Invalid: UT & 9 &	14.40 &	7.20 &	12.00 &	20.80\\		
			Valid \& Invalid: UT & 46 &	23.20 &	16.90 &	21.10 &	47.50\\		
			Valid: IT & 50 &	70.00 &	61.40 &	80.10 &	85.10\\		
			Invalid: IT & 44 &	76.80 &	67.50 &	77.20 &	77.20\\		
			Valid \& Invalid: IT & 94 &	90.10 &	86.70 &	94.90 &	91.10\\		
			Valid: UT \& IT & 87 &	70.80 &	62.00 &	80.90 &	88.10\\		
			Invalid: UT \& IT & 53 &	78.30 &	71.10 &	78.30 &	81.20\\		
			Valid \& Invalid: UT \& IT & 140 &	91.20 &	90.40 &	95.70 &	94.10\\	
			\hline
		\end{tabular}
		}
		\caption{Coverage metrics measured for varying runs throughout the experiment}
		\label{table:Coverage Metrics}
	\end{center}
\end{table}

The first column displays the combination of unit (UT), integration (IT), valid and invalid request handling tests considered for each run. 
The second column presents the number of tests executed for each run. 
Finally, the remaining columns represent various coverage metrics including: instruction, branch, line and method coverage. 

\subsubsection{Evaluating Correctness of inferred Specifications}
\label{subsubsection:EvalCorrectnessSpec}

Deciding whether the inferred invariants for each run were correct or incorrect required some thought. 

\textbf{Correct invariants: } Consider a \texttt{deposit} method that accepts an \textit{amount} value to be added to an existing \textit{balance}, as shown in Listing \ref{listing:DepositMethodExample}. 
A property that could be defined for such a function would be that the \texttt{amount} variable upon execution of the \textit{deposit} method has to be greater or equal to zero. 

\begin{lstlisting}[caption=Snippet code of a deposit function, label=listing:DepositMethodExample]
public class UserAccount {
   protected boolean opened;
   protected String account_number;
   protected double balance;
   ...	

   public void deposit(double amount) 
   {
      balance += amount;
   }
   ...
\end{lstlisting}

\begin{lstlisting}[caption=Snippet code of a deposit test function, label=listing:DepositTestExample]
public class UserAccountUTPass {

   Integer userId = 03;
   String accountNumber = "45678";
   double startingBalance = 35.0;
   double amountToDeposit = 100.0;
   ...


   @Test
   public void deposit_Test_Pass(){
      userAccount.balance = startingBalance;
      userAccount.deposit(amountToDeposit);
      assertTrue(userAccount.balance == (startingBalance + amountToDeposit));
   }
   ...
\end{lstlisting}

An invariant inferred by the tool when exercising the software using the test shown in Listing \ref{listing:DepositTestExample} that would comply with the latter property, would be: 

\begin{center}
	amount \textgreater= 35.0
\end{center}

However, as one would notice, the inferred invariant does not fully represent the set of all acceptable input since it would not accept values between zero and thirty four. 
Such an invariant is deemed to over approximate, limiting the range of acceptable input. 
Similarly, there can be under approximating invariants that include the set of acceptable input whilst also accepting some undesirable ones \cite{OverUnderApprox}. 
Since the study wanted to determine to what extent the inference technology was able to infer complete specifications, conditions were therefore classified as correct only if they did not over or under approximate at least one property inside the manually written specification. 
 
Throughout the evaluation process however, some inferred conditions were encountered that correctly expressed how the system should behave but did not coincide with any property from the manually written specification. 
In such instances, the observed conditions were deemed to be correct and added to the manually written specification, expecting the other runs to also infer the newly added condition.

\textbf{Incorrect Invariants: } The remaining conditions inside the inferred specifications were then classified as incorrect. 

\textbf{Missing Invariants: } Having classified all the inferred conditions, the properties residing inside the manually written specification that did not have a matching invariant were marked as missed. 

The resulting classifications for each run can be found in Table \ref{table:PrecRecallResults}.


\begin{table}[H]
	\centering
	\begin{adjustbox}{max width=\textwidth}
	\begin{tabular}{cccccccccclll}
		\cline{7-10}
		&                                     &                                                      &                                                        &                                   & \multicolumn{1}{c|}{}                               & \multicolumn{2}{c|}{Original Spec}                                   & \multicolumn{2}{c|}{\cellcolor[HTML]{C0C0C0}Original + Extra Spec}                                                   &  &  &  \\ \cline{1-10}
		\multicolumn{1}{|c|}{Runs}                       & \multicolumn{1}{c|}{Correct (Orig)} & \multicolumn{1}{c|}{\cellcolor[HTML]{C0C0C0}Correct} & \multicolumn{1}{c|}{\cellcolor[HTML]{C0C0C0}Incorrect} & \multicolumn{1}{c|}{Missed(Orig)} & \multicolumn{1}{c|}{\cellcolor[HTML]{C0C0C0}Missed} & \multicolumn{1}{c|}{Precision(\%)} & \multicolumn{1}{c|}{Recall(\%)} & \multicolumn{1}{c|}{\cellcolor[HTML]{C0C0C0}Precision(\%)} & \multicolumn{1}{c|}{\cellcolor[HTML]{C0C0C0}Recall(\%)} &  &  &  \\ \cline{1-10}
		\multicolumn{1}{|c|}{Valid: UT}                  & \multicolumn{1}{c|}{448}            & \multicolumn{1}{c|}{\cellcolor[HTML]{C0C0C0}473}     & \multicolumn{1}{c|}{\cellcolor[HTML]{C0C0C0}982}       & \multicolumn{1}{c|}{469}          & \multicolumn{1}{c|}{\cellcolor[HTML]{C0C0C0}755}    & \multicolumn{1}{c|}{31.32}         & \multicolumn{1}{c|}{48.85}      & \multicolumn{1}{c|}{\cellcolor[HTML]{C0C0C0}32.51}         & \multicolumn{1}{c|}{\cellcolor[HTML]{C0C0C0}38.52}      &  &  &  \\
		\multicolumn{1}{|c|}{Invalid: UT}                & \multicolumn{1}{c|}{188}            & \multicolumn{1}{c|}{\cellcolor[HTML]{C0C0C0}199}     & \multicolumn{1}{c|}{\cellcolor[HTML]{C0C0C0}517}       & \multicolumn{1}{c|}{729}          & \multicolumn{1}{c|}{\cellcolor[HTML]{C0C0C0}1029}   & \multicolumn{1}{c|}{26.67}         & \multicolumn{1}{c|}{20.50}      & \multicolumn{1}{c|}{\cellcolor[HTML]{C0C0C0}27.80}         & \multicolumn{1}{c|}{\cellcolor[HTML]{C0C0C0}16.21}      &  &  &  \\
		\multicolumn{1}{|c|}{Valid \& Invalid: UT}       & \multicolumn{1}{c|}{481}            & \multicolumn{1}{c|}{\cellcolor[HTML]{C0C0C0}507}     & \multicolumn{1}{c|}{\cellcolor[HTML]{C0C0C0}1032}      & \multicolumn{1}{c|}{436}          & \multicolumn{1}{c|}{\cellcolor[HTML]{C0C0C0}721}    & \multicolumn{1}{c|}{31.79}         & \multicolumn{1}{c|}{52.45}      & \multicolumn{1}{c|}{\cellcolor[HTML]{C0C0C0}32.94}         & \multicolumn{1}{c|}{\cellcolor[HTML]{C0C0C0}41.29}      &  &  &  \\
		\multicolumn{1}{|c|}{Valid: IT}                  & \multicolumn{1}{c|}{854}            & \multicolumn{1}{c|}{\cellcolor[HTML]{C0C0C0}903}     & \multicolumn{1}{c|}{\cellcolor[HTML]{C0C0C0}2028}      & \multicolumn{1}{c|}{63}           & \multicolumn{1}{c|}{\cellcolor[HTML]{C0C0C0}325}    & \multicolumn{1}{c|}{29.63}         & \multicolumn{1}{c|}{93.13}      & \multicolumn{1}{c|}{\cellcolor[HTML]{C0C0C0}30.81}         & \multicolumn{1}{c|}{\cellcolor[HTML]{C0C0C0}73.53}      &  &  &  \\
		\multicolumn{1}{|c|}{Invalid: IT}                & \multicolumn{1}{c|}{604}            & \multicolumn{1}{c|}{\cellcolor[HTML]{C0C0C0}658}     & \multicolumn{1}{c|}{\cellcolor[HTML]{C0C0C0}1627}      & \multicolumn{1}{c|}{313}          & \multicolumn{1}{c|}{\cellcolor[HTML]{C0C0C0}570}    & \multicolumn{1}{c|}{27.07}         & \multicolumn{1}{c|}{65.87}      & \multicolumn{1}{c|}{\cellcolor[HTML]{C0C0C0}28.80}         & \multicolumn{1}{c|}{\cellcolor[HTML]{C0C0C0}53.58}      &  &  &  \\
		\multicolumn{1}{|c|}{Valid \& Invalid: IT}       & \multicolumn{1}{c|}{878}            & \multicolumn{1}{c|}{\cellcolor[HTML]{C0C0C0}934}     & \multicolumn{1}{c|}{\cellcolor[HTML]{C0C0C0}2294}      & \multicolumn{1}{c|}{39}           & \multicolumn{1}{c|}{\cellcolor[HTML]{C0C0C0}294}    & \multicolumn{1}{c|}{27.68}         & \multicolumn{1}{c|}{95.75}      & \multicolumn{1}{c|}{\cellcolor[HTML]{C0C0C0}28.93}         & \multicolumn{1}{c|}{\cellcolor[HTML]{C0C0C0}76.06}      &  &  &  \\
		\multicolumn{1}{|c|}{Valid: UT \& IT}            & \multicolumn{1}{c|}{863}            & \multicolumn{1}{c|}{\cellcolor[HTML]{C0C0C0}912}     & \multicolumn{1}{c|}{\cellcolor[HTML]{C0C0C0}2011}      & \multicolumn{1}{c|}{54}           & \multicolumn{1}{c|}{\cellcolor[HTML]{C0C0C0}316}    & \multicolumn{1}{c|}{30.03}         & \multicolumn{1}{c|}{94.11}      & \multicolumn{1}{c|}{\cellcolor[HTML]{C0C0C0}31.20}         & \multicolumn{1}{c|}{\cellcolor[HTML]{C0C0C0}74.27}      &  &  &  \\
		\multicolumn{1}{|c|}{Invalid: UT \& IT}          & \multicolumn{1}{c|}{673}            & \multicolumn{1}{c|}{\cellcolor[HTML]{C0C0C0}727}     & \multicolumn{1}{c|}{\cellcolor[HTML]{C0C0C0}1829}      & \multicolumn{1}{c|}{244}          & \multicolumn{1}{c|}{\cellcolor[HTML]{C0C0C0}501}    & \multicolumn{1}{c|}{26.90}         & \multicolumn{1}{c|}{73.39}      & \multicolumn{1}{c|}{\cellcolor[HTML]{C0C0C0}28.44}         & \multicolumn{1}{c|}{\cellcolor[HTML]{C0C0C0}59.20}      &  &  &  \\
		\multicolumn{1}{|c|}{Valid \& Invalid: UT \& IT} & \multicolumn{1}{c|}{898}            & \multicolumn{1}{c|}{\cellcolor[HTML]{C0C0C0}954}     & \multicolumn{1}{c|}{\cellcolor[HTML]{C0C0C0}2340}      & \multicolumn{1}{c|}{19}           & \multicolumn{1}{c|}{\cellcolor[HTML]{C0C0C0}274}    & \multicolumn{1}{c|}{27.73}         & \multicolumn{1}{c|}{97.93}      & \multicolumn{1}{c|}{\cellcolor[HTML]{C0C0C0}28.96}         & \multicolumn{1}{c|}{\cellcolor[HTML]{C0C0C0}77.69}      &  &  &  \\ \cline{1-10}
		\multicolumn{1}{l}{}                             & \multicolumn{1}{l}{}                & \multicolumn{1}{l}{}                                 & \multicolumn{1}{l}{}                                   & \multicolumn{1}{l}{}              & \multicolumn{1}{l}{}                                & \multicolumn{1}{l}{}               & \multicolumn{1}{l}{}            & \multicolumn{1}{l}{}                                       & \multicolumn{1}{l}{}                                    &  &  & 
	\end{tabular}
	\end{adjustbox}
	\caption{Classification results for multiple runs including precision and recall calculations}
	\label{table:PrecRecallResults}
\end{table}

The first column shows the combination of unit, integration, valid and invalid request testing that were considered for each run. 
The second column represents the number of inferred invariants observed to correctly represent properties defined in the manually written specification, whilst the fifth column, shows the number of properties defined in the specification that were not represented. 
The greyed third, fourth and sixth columns, present the total number of correct (including the extra invariants added to the original expected set), incorrect and missed (including those missed from the extra invariants) invariants that were observed. 
Finally, the last four columns present the precision and recall of the inferred invariants based on the original reported findings and the combination of original with the extra invariants(grey columns) respectively. 

The difference between the varying runs can be clearly seen by the number of correct versus missing conditions presented. 
Runs that only considered unit tests faired the worst, observing a higher number of missing conditions than correct ones. 
On the other hand, runs that made use of integration tests performed better, some even providing over 900 correct conditions. 
The last run that combined all unit and integration tests faired the best, exhibiting the highest number of correct conditions observed at 954 versus 274 missed for the combined results and 898 correct versus 19 missed for the original spec comparison. 
The recall column illustrates this, registering an increase in quality of over 77.43 and 61.48 percentage points for the original and combined results over the worst performance of 20.50 and 16.2 percent at 97.93 and 77.69 percent respectively.  
However, the precision column did not present similar increases. 
Instead, the precision percentage was observed to decrease as the complexity of the runs increased. 
This occurred since precision takes into consideration the number of incorrect invariants, whom kept increasing as the coverage values increased.
Moreover, recall percentages were observed to be higher when taking into account the classification reports that do not include the extra invariants. 

\subsubsection{Discovering test suite attributes that influence quality of inferred specification}
\label{subsubsection:DiscAttQual}

Once all the metrics for each run were measured and invariants classified, the study to discover correlations between high quality specification inference and test suite attributes was carried out. 

The precision and recall values along with the coverage metrics presented in Tables \ref{table:PrecRecallResults} and \ref{table:Coverage Metrics} were passed on to a linear estimation function to identify correlations between the two. 
As expected, instruction, method and branch coverage attributes were found to have a high correlation to the quality of the inferred specification with an 0.8956 determination coefficient for recall.
This was as (discussed in Section \ref{subsection:ChoosingRelevantAttributes}) expected since such high coverage values indicate that a greater chunk of the software was exercised, increasing the behaviour observed and chances of inferring useful invariants 

The linear estimation function found a clear correlation, however, having high instruction, method and branch coverage does not ensure that the inference tool will produce the optimal set of invariants. 
In fact, although the number of correct conditions observed (for the optimal run) was considerable, many of the correct invariants did not represent the manually written specifications. 
Instead, most consisted of additional invariants that had been inferred by Daikon and added throughout the classification process since they were found to be correct and useful. 
Furthermore, it was observed that some properties defining major functionality of the software were never observed, highlighting a number of deficiencies and limitations. 

One limitation that was observed whilst identifying the missed conditions, was the ability to reason about complex data types such as \textit{ENUMS} and array lists of complex data types. 
Hence, a sizeable amount of the specification was never met. 
Moreover, the Daikon tool was observed to ignore the inference of invalid input handling whenever exceptions were thrown by the system.  

Other complex properties such as

\begin{center}
	\textit{ Return true only if the User ID is valid, Session ID is open, account\_number exists, status of account is set to open, the amount is greater than zero and there are sufficient funds for a withdrawal request }
\end{center}

were not observed. 
However, this was expected since such properties involve the combination of more than three variables for one invariant, which is one major limitation that the Daikon tool has. 
Moreover, for a tool to infer such a property, some level of knowledge with regards to what the data represents would be required which is hard to obtain without requiring human aid. 

Throughout the classification process, there were also some conditions which were expected but never observed including:

\begin{center}
	
	orig(balance) + amount == balance
	
	uid \textgreater= 0
	
	amount \textgreater= 0
\end{center}

Having a property checking that the balance is equal to the original balance plus the amount, or checking that the user id (\textit{uid}) and \textit{amount} is greater than zero are all examples of invariants that the Daikon tool is known to be capable of inferring. 
However, they did not appear in any of the nine runs evaluated. 
Upon further investigation, it was discovered that the Daikon tool requires a more diverse test set for such properties to be inferred. 
This was confirmed since incorrect conditions like ``amount one of \{200,100,50\}'' were observed confirming that the tool was not able to further generalise the invariant and required more diverse values to do so. 

Therefore, having high instruction, method and branch coverage does not ensure that the inference tool will produce the optimal set of specifications. 
Nonetheless, they contribute to the quality of inferred specifications. 

\subsection{Discussion: Lessons learned}

We have discovered through the tool agnostic experiment presented in this paper that instruction, method and branch coverage metrics influence the quality of specifications inferred. 
However, they do not ensure that the specifications inferred will be correct. 

Given the observations that we made, using fixed values to test similar behaviour is not recommended. 
Instead, a wider input selection should be used to increase the chances of inferring better specifications. 
However, due to the inference nature of tools such as Daikon, using a wider range of input may not be enough. 
The additional inputs distributed across tests validating different properties, would still provide limited samples ending up with inaccurate invariants. 
Therefore, in addition to using a wider rage of input, we would also recommend to increase the number of times the software's functionality is exercised using unique inputs each time. 
Hence providing more samples for the same behaviour. 

Taking an equivalent partitioning approach and running the test suite multiple times may also increase the chances of inferring better quality specifications since a potentially wide selection of inputs for each corner case and branch opportunity would be available providing multiple unique samples from which specifications can be correctly inferred. 

By providing more samples, tools such as Daikon, will be able to refine the invariant generalisation process and potentially increase the number of correct invariants.  
Moreover, the increase in unique samples may also drastically reduce the number of incorrect invariants since the chances of having an incorrect invariant invalidated by a sample during the inference process also increases.

	\section{Related Work}
\label{section:RelatedWork}

Several studies have been conducted to determine whether test suite attributes influence the quality of specifications inferred. 
Harder, Mellen and Ernst \cite{ImpTestSuitesViaOpAbst}, presented a new metric that can be used to write, generate and minimise test suites. 
To evaluate this new approach, various existing metrics including statement and branch coverage were taken into consideration and compared. 
The comparison was carried out by instructing Daikon to infer invariants for faulty and correct versions of C programs driven by test suites written to obtain high respective coverage values. 
Differences in effectiveness were obtained by ``measuring the proportion of times the fault was detected'' by  \cite{ImpTestSuitesViaOpAbst} each test suite. 
For our purpose, the quality of invariants inferred using varying test suites cannot be measured by counting the number of faults detected. 
This is due to the fact that our end goal is that to obtain a specification.
Therefore, it is of the utmost importance that the system operates as intended without any faults to be able to correctly infer the specification. 
Nonetheless, the same conclusions were drawn.
Having 100\% statement and branch coverage was confirmed to aid fault finding. 
However, if other metrics such as the one introduced by their study is combined, further improvements can be observed. 
This conforms with the observations made by this study that using other metrics may potentially refine and produce more accurate invariants. 

Similarly, Gupta and Heidepriem \cite{Gupta}, introduced the \textit{invariant-coverage} criterion aimed to improve the accuracy of dynamically inferred specifications. 
The authors made use of five programs covering a variety of program features such as calculating the Levenshtein distance between two strings and calculating the adjoint of a matrix. 
Throughout the evaluation process, the specifications inferred when using traditional branch and definition-use pair suites were compared to the outcome of the newly introduced criterion. 
Although only traditional coverage metrics were taken into consideration for our study, a larger program was used to challenge both the abilities of the specification inference tool and the test suites used to infer the specifications.
Moreover, a different approach was taken to compare the effectiveness of using different coverage metrics.  
Instead of comparing the inferred invariants to a formally written specification, the authors identified meaningful invariants that are not ''linked to specific input values used and are abstractions over multiple inputs''. 
Nevertheless, Gupta et al. showed that \textit{invariant-coverage} suites were able to infer higher quality specifications eliminating various false invariants inferred when using traditional criteria. 
Confirming yet again the results of the study that we carried out.

Polikarpova, Ciupa and Meyer \cite{CompStudyProgWrittenAutInfCont}, performed experiments to compare properties that are written by developers and testers versus those inferred automatically by tools such as Daikon. 
Much like the experiment that was carried out by this paper, the authors measured the proportion of user written properties implied by the inferred specification and the number of incorrect invariants inferred, finding correlations between code metrics and the quality of contracts inferred. 
The total recall was that of 59\% for the large test suite whilst negative correlations were observed between correctness and coupling. 
Positive correlations on the other hand were observed between assertion clauses inferred and the number of relevant inferred invariants.

\section{Conclusion}
\label{section:Conclusion}

No one doubts the importance of formally written specifications. 
However, such specifications are rarely written out and when they are, checking their correctness and completeness is hard. 
Instead of forcing developers or testers to write such specifications, various tools including Daikon have explored the possibility of automatically inferring generalised specifications based on sources of information that are provided including test suites. 
Given the nature of such sources, the pool of information may be limited, which in turn reflects negatively upon the outcome of the specifications inferred. 

This study carried out an experiment to discover test suite attributes that positively influence the quality of the inferred specifications. 
Results indicated that there was a strong correlation between having high line, method and branch coverage and high recall values (up to 97.93 \%). 
However, even though the quality of the inferred specifications was high, it was also observed that having high coverage values does not ensure that optimal specifications will be inferred. 
As a consequence, incorrect and missing invariants were still observed highlighting the importance of having a varied selection of inputs from which such techniques can correctly generalise the expected behavioural conditions.
In fact, this study can be applied to other specification inference tools whereby the same or similar results are expected. 
	
	\subsection{Future Work}
\label{section:FutureWork}

This paper presents precision and recall values for specifications inferred by Daikon when provided with varying test suites. 
As discussed in Section \ref{subsubsection:DiscAttQual}, in addition to having high instruction and branch coverage, a wider selection of inputs and exercises are required for Daikon to be able to successfully infer correct specifications. 
In future work, we aim to investigate other possible approaches that can be used to automatically introduce such inputs and drive through the use of techniques such as random and concolic testing aming to obtain the optimal specification. 
Metrics presented by Gupta et al. \cite{Gupta} and Harder et al. \cite{ImpTestSuitesViaOpAbst} could also be used alongside such techniques to enhance the quality of specifications inferred. 

Furthermore, after concluding the experiments on the small case study, we would like to carry out a similar test on a larger industrial system to evaluate whether the industry is ready to adopt the technology and provide guidelines on how the test suites should be written to ensure optimal specification inference results. 

Iterative feedback loops \cite{FeedDrivDynInvDisc} are also of interest to us since the existing test suites could potentially be used as the initial ``seeds'' from which the process can start inferring a specification and gradually refine it every iteration.
Moreover, we are not excluding the possibility of exploring other techniques that do not use Daikon such a DySy \cite{DySy} which combines concrete executions of test suites with symbolic execution to abstract conditions over program variables satisfied by tests during execution. 

Finally, we would also like to explore other possible sources of information that could be introduced to the trace recording phase providing a richer understanding of how the system behaves. 
An example of sources that we are considering include live system runs or execution logs. 
By taking such sources into consideration, we are hypothesising the possibility that the number of incorrect invariants classified would decrease drastically since many of the incorrect invariants would be invalidated by traces that contradict them. 
In addition, some of the partially correct invariants could also potentially mutate into correct invariants given the broader set of behaviour observed.

	\nocite{*}
	
	\bibliographystyle{eptcs}
	\bibliography{generic}
	
\end{document}